\documentclass[acmsmall,natbib=false]{acmart}
\setcopyright{none}
\renewcommand\footnotetextcopyrightpermission[1]{}
\usepackage{hyperref}
\usepackage{caption}
\usepackage{subcaption} 
\usepackage[most]{tcolorbox}
\usepackage[linesnumbered,ruled,vlined]{algorithm2e}

\tcbset{textmarker/.style={%
        enhanced,
        parbox=false,boxrule=0mm,boxsep=0mm,arc=0mm,
        outer arc=0mm,left=3mm,right=3mm,top=7pt,bottom=7pt,
        toptitle=1mm,bottomtitle=1mm,oversize}}

\usepackage[most]{tcolorbox}
\tcbset{textmarker/.style={%
        enhanced,       parbox=false,boxrule=0mm,boxsep=0mm,arc=0mm,
        outer arc=0mm,left=3mm,right=3mm,top=7pt,bottom=7pt, toptitle=1mm,bottomtitle=1mm,oversize}}

\newcounter{obscounter}
\renewcommand{\theobscounter}{\Roman{obscounter}}
\definecolor{custompurple}{HTML}{53257F}
\newtcolorbox{ObservationBox}[2][]{text width=0.96\linewidth,
colbacktitle=custompurple,enhanced,
attach boxed title to top left={yshift=-2mm,xshift=3mm},
boxed title style={sharp corners},top=6pt,bottom=2pt,title=#2,colback=custompurple!10!white, left=4pt, right=2pt}
\newcommand{\obs}[1]{
\refstepcounter{obscounter}
\begin{ObservationBox}{\textbf{Observation \theobscounter}}
\par\noindent
#1
\end{ObservationBox}}

\RequirePackage[
  datamodel=acmdatamodel,
  style=acmnumeric,
  ]{biblatex}

\begin{document}
\pagestyle{empty}
%%
%% The "title" command has an optional parameter,
%% allowing the author to define a "short title" to be used in page headers.
\title{Taking Cryptography Out of the Data Path via Near-Memory Processing in DRAM}

\author{Nicola Barcarolo}
%\authornote{Both authors contributed equally to this research.}
\email{nicola.barcarolo@studenti.unitn.it}
%\orcid{xxxx-xxxx-xxxx}
%\authornotemark[1]
\affiliation{%
  \institution{University of Trento}
  \country{Italy}
}

\author{Brahmaiah Gandham}
% \authornotemark[1]
\affiliation{%
 \institution{University of Trento}
 \country{Italy}}
 \email{brahmaiah.gandham@unitn.it}

\author{Mohammad Sadrosadati}
\affiliation{%
  \institution{ETH Zurich}
  %\city{Haidian Qu}
  %\state{Beijing Shi}
  \country{Switzerland}}
  \email{m.sadr89@gmail.com}

\author{Roberto Passerone}
\affiliation{%
  \institution{University of Trento}
  %\city{San Antonio}
  %\state{Texas}
  \country{Italy}}
\email{roberto.passerone@unitn.it}

\author{Onur Mutlu}
\affiliation{%
  \institution{ETH Zurich}
  %\city{Hekla}
  \country{Switzerland}}
\email{omutlu@gmail.com}

\author{Flavio Vella}
\affiliation{%
  \institution{University of Trento}
  %\city{I}
  \country{Italy}}
\email{flavio.vella@unitn.it}

%%
%% By default, the full list of authors will be used in the page
%% headers. Often, this list is too long, and will overlap
%% other information printed in the page headers. This command allows
%% the author to define a more concise list
%% of authors' names for this purpose.
\renewcommand{\shortauthors}{Barcarolo et al.}

\begin{abstract}
  %Abstract a bit of Context challenge and results
Cryptographic algorithms such as AES-128 and SHA-256 are fundamental to ensuring data security and integrity. Although these algorithms are computationally efficient, their performance is often constrained by the processor-centric architectures (e.g., CPUs, GPUs), primarily due to the memory bottleneck. This constraint leads to increased latency and higher energy consumption, particularly when handling large volumes of data. To overcome these challenges, Processing-in-Memory (PIM) has emerged as a promising architectural paradigm, allowing computation to occur directly within or near memory units. By minimizing data movement between the processor and memory units, PIM can significantly accelerate cryptographic algorithms while improving energy efficiency. Several pieces of prior work have demonstrated the effectiveness of PIM at fundamentally accelerating cryptographic algorithms. However, none of the prior works have extensively demonstrated the potential of a real-world PIM system. In this paper, we want to investigate the potential and limitations of real-world PIM in accelerating cryptographic algorithms. As part of our methodology, the UPMEM PIM architecture is used to assess the scalability of cryptographic algorithms. When these algorithms operate on a single rank, their performance remains below that of modern CPUs. However, distributing the computation across multiple ranks significantly enhances performance. When all available ranks are utilized, real-world PIM can accelerate cryptographic algorithms more effectively.
\end{abstract}

% \footnotetext[1]{New Paper, Not an Extension of a Conference Paper.}

%%
%% The code below is generated by the tool at http://dl.acm.org/ccs.cfm.
%% Please copy and paste the code instead of the example below.
%%

%%
%% Keywords. The author(s) should pick words that accurately describe
%% the work being presented. Separate the keywords with commas.

% \received{20 February 2007}
% \received[revised]{12 March 2009}
% \received[accepted]{5 June 2009}

%%
%% This command processes the author and affiliation and title
%% information and builds the first part of the formatted document.
\maketitle
\thispagestyle{empty}

%% The next two lines define the bibliography style to be used, and
%% the bibliography file.
%\footnotetext[1]{Footnote text}
%\footnote{New Paper, Not an Extension of a Conference Paper.}
\section{Introduction}
% to review
The Advanced Encryption Standard (\emph{AES})~\cite{AES_FIPS} and the Secure Hash Algorithm (\emph{SHA})~\cite{NIST_SHA} are two of the most widely used algorithms where data security is needed. However, their software implementation can be computationally expensive and can provide only a limited level of throughput due to the \emph{memory-bound} characteristics of the algorithms. In systems based on the processor-centric architecture, the performance of some tasks is limited not by the CPU computational power, but by the bottleneck between the CPU and main memory. This bottleneck is caused by high-latency memory access and data transfers over a narrow-bandwidth communication channel. Furthermore, this bottleneck is responsible for a significant portion of total energy consumption. Recent studies show that data movement between the processor and RAM can account for up to 62\% of total system energy consumption~\cite{PrimerPIM}.

To address the limitations of memory bandwidth in processor-centric architectures, a growing number of \emph{Processing-In-Memory} (\emph{PIM}) architectures have been proposed in recent years~\cite{pim1,PrimerPIM,pim3}, attracting increase interest from both the academic community~\cite{p1,p2,p3,p4,p5,p6,p7,p8,p9,p10,p11,p12,p13,p14,p15,p16,PIM_WLProspective,p18,p19,p21,p22,p24,he,p27,p28,p29,p30,p31,p32,p33,p34,p35,p36,p37,p38,p39,p40,pim1,p42,p45,p46,p47,p48,pim3,p50,p51,p52,p53,p54,p61,p56} and industry. This renewed momentum is reflected in the emergence of several commercial systems and hardware prototypes that integrate PIM capabilities~\cite{p64,p65,FIMDRAM_2,p67,p68,p69,p70,p71}.

\begin{figure}[t]
\centering
\includegraphics[scale=0.75]{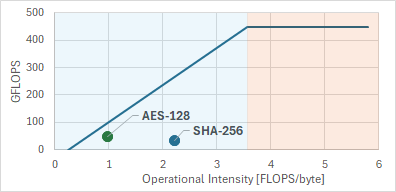}
\caption{Roofline model applied on the UPMEM processor-centric architecture: only standard CPU (\emph{INTEL Xeon Silver 4215}) and non-PIM memory (\emph{DDR4-2666 RDIMM Dual Rank DRAM}) is considered}
\label{fig:Roofline}
\end{figure}

The core idea behind \emph{PIM} is to integrate compute-capable units directly within the memory chips, enabling data to be processed in place and thus avoiding the performance and energy penalties associated with transferring data over a narrow-bandwidth channel.
Due to technological constraints, the practical realization of such architectures has only become feasible recently. UPMEM was the first to design and produce a working \emph{Processing-In-Memory} system, the details of which are discussed in Section~\ref{sec:Background}. The software implementation of both AES-128 and SHA-256 has \emph{memory-bound} characteristics. Figure~\ref{fig:Roofline} shows that both algorithms are placed in the memory-bound region of the Roofline model~\cite{Roofline} when applied to the UPMEM non-PIM architecture (i.e., a processor-centric architecture, see Section~\ref{sec:Background}).

%Keep?
%The horizontal line of the roofline model takes into account the theoretical CPU peak performance (\emph{P}), which depends on three main factors: 1)The CPU clock speed, 2)The number of processing units in the architecture (i.e. number of CPU cores), 3)The number of FLOPS per cycle executed by the CPU, if it supports SIMD units (for example using \emph{Advanced Vector Extensions}, a set of ISA extensions that enables SIMD instructions on CPU).
%The slope of the oblique line, instead, takes into consideration the performance limitation due to memory bandwidth (\emph{B}). This line can be analytically expressed by the multiplication between memory bandwidth \emph{B} and \emph{OI}.
Despite real-world PIM systems have been widely tested with workloads defined as memory-bound from very different domains (algebra~\cite{p18}, bio-informatics\cite{p62,p35}, image processing~\cite{p63}, neural networks~\cite{p58}, etc.)~\cite{BenchRealPIM_UPMEM}, among cryptographic algorithms, only one study has addressed AES-128~\cite{PIMencryptionSASHA}, and the present work is the first to implement SHA-256.\\

This paper makes several key contributions to the PIM computing field:
\begin{itemize}
    \item It provides a practical example of how to design and test a PIM-enabled program on a real-world PIM system, showcasing its real-world capability.
    \item It characterizes the performance of the two most widely-used encryption algorithms, \emph{AES-128} and \emph{SHA-256}, on a machine adopting a \emph{Processing-Near-Memory} approach, highlighting how these memory-bound algorithms behave under a real-world PIM system.
    \item It explores the potential and limitations of PIM machines to accelerate the \emph{AES-128} and \emph{SHA-256} algorithms, investigating whether and how PIM systems can be exploited to enhance their speed-up compared to processor-centric architectures.
\end{itemize}

This paper is organized as follows. At the beginning, Section~\ref{sec:Background} provides a brief overview of the main \emph{Processing-In-Memory} approaches, along with a description of the UPMEM architecture and its key features. Section~\ref{sec:relatedwork} reviews prior research on cryptographic workloads in PIM systems. Section~\ref{sec:cryptoPIM} introduces a software-based PIM implementation of \emph{AES-128} and \emph{SHA-256}, while Section~\ref{sec:results} presents the performance results and compares them with those of traditional software implementations and, in the case of \emph{AES-128}, also with a hardware implementation.

%\flv{Add the organization of the paper: what are the roles of each section. }

\section{Background} \label{sec:Background}
This section presents the first publicly available real-world PIM system, with particular emphasis on its main features, programming model, and a brief characterization of its achievable performance. Although the first studies on \emph{PIM} date back over fifty years, the implementation of a memory-centric architecture remained a major challenge until the last decade, primarily due to technological limitations in RAM fabrication~\cite{PrimerPIM}. In recent times, DRAM scaling has reached its practical limits, and it has become very difficult to improve density, reduce latency, and lower energy consumption~\cite{PIM_WLProspective}. In response, designers have started exploring alternative solutions by developing new memory devices based on emerging technologies, thereby paving the way for novel memory-centric architectures.

In recent years, two major approaches have been developed for \emph{PIM} architectures. The first approach is \emph{Processing-Using-Memory} (\emph{PUM}), which leverages the intrinsic properties of analog signals to perform computations directly within memory chips. For example, many \emph{PUM} architectures utilize modified sense amplifiers that enable the simultaneous activation of three memory rows. This approach requires only minimal modifications compared to standard RAM structures. However, in most cases, only simple operations, such as addition or bit-wise logical operations, can be executed within the memory. An alternative approach is \emph{Processing-Near-Memory} (\emph{PNM}), as implemented by the company UPMEM, which involves integrating compute-capable units, such as accelerators or general-purpose cores, directly inside or in close proximity to memory chips. This design enables memory to play an active role in data processing, significantly avoiding the need for extensive data transfers over a narrow-band bus~\cite{PrimerPIM}.

\subsection{The UPMEM architecture}
\begin{figure}[t]
\begin{center}
    \includegraphics[scale=0.38]{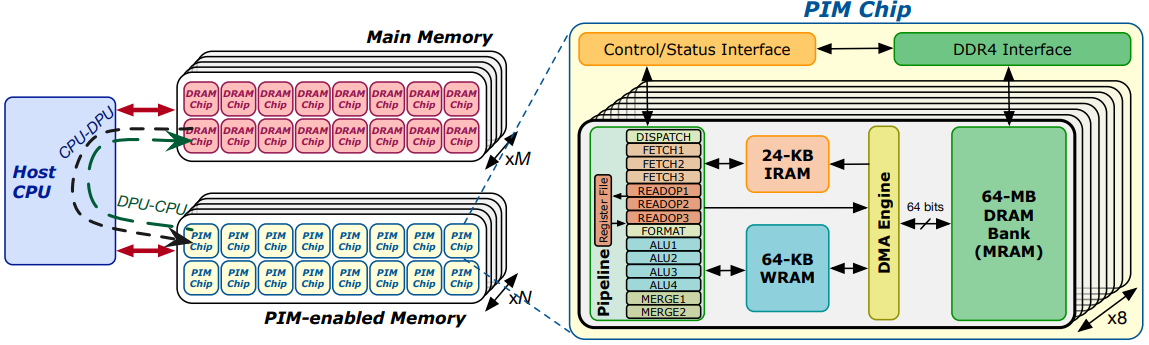}
    \captionof{figure}{High-level view of UPMEM architecture~\cite{BenchRealPIM_UPMEM}.\\}
    \label{fig:UPMEM_Architecture}
\end{center}
\end{figure}

A high-level representation of the UPMEM PIM architecture~\cite{UPMEMbench} is shown in Figure~\ref{fig:UPMEM_Architecture}. The main memory includes both traditional DRAM and \emph{PIM-enabled} DDR4 DRAM, which integrates general-purpose cores, known as \emph{DRAM Processing Units} (\emph{DPUs}), directly within its structure.

Each memory Dual In-line Memory Module (DIMM) contains eight or sixteen PIM chips, and each chip is equipped with eight \emph{DPUs}, which are multi-threaded 32-bit RISC cores featuring a dedicated instruction set architecture (\emph{ISA})~\cite{BenchRealPIM_UPMEM}. Each DPU features a 13-stage pipeline capable of simultaneously handling up to 24 \emph{tasklets}, which are hardware threads that share the same memory space for both code and data, with each tasklet able to execute different instructions from the same program.

Each DPU has access to a 24~kB SRAM instruction memory (IRAM), which stores the instructions to be executed by the DPU, and to a 64~kB SRAM working memory (WRAM), used as scratchpad memory. Both are shared among all \emph{tasklets} running on the DPU~\cite{BenchRealPIM_UPMEM}. In addition, each DPU has access to a 64~MB main memory (MRAM), which can also be accessed by the host CPU to transfer data to or from the system main memory~\cite{UPMEMbench}~\cite{RealPIM}.

\subsection{Programming model}
The programming model adopted by UPMEM is the \emph{Single Program Multiple Data} (\emph{SPMD}), in which all DPUs execute the same set of instructions but operate on different data~\cite{BenchRealPIM_UPMEM}. The main CPU (a standard \emph{Intel Xeon Silver 4215} CPU) manages the allocation, execution, and control of the DPUs. In the host application, it is possible to allocate an arbitrary number of ranks, each consisting of 64 DPUs, or to allocate a specific number of DPUs. Subsequently, the same DPU program, which must be a specially pre-compiled version, is loaded onto all allocated DPUs. The host application also manages data transfers between the CPU and DPUs in both directions, as well as the launching of DPUs. These operations can be performed in either synchronous or asynchronous mode. In synchronous mode, the host CPU thread is suspended until the DPUs complete kernel execution, whereas in asynchronous mode, control is returned to the host CPU thread immediately~\cite{BenchRealPIM_UPMEM}. Once the data has been transferred to the DPUs, they can begin execution in parallel. Because each DPU and its memory are integrated on the same chip, the DRAM Processing Units can perform computations without needing to transfer data through a narrow-band channel, as is required in traditional processor-centric architectures.

Furthermore, each DPU program can utilize an arbitrary number of \emph{tasklets}, up to a maximum of 24, which must be specified at compile time. Each \emph{tasklet} functions as a thread that shares the instruction memory (IRAM) with the others. The data memory, including both WRAM and MRAM, is also shared among all \emph{tasklets}. However, tasklets can allocate private variables by declaring them as local, for example, within the main function. Due to the shared-memory model, synchronization mechanisms may be required to manage critical sections. To support this, the UPMEM programming model provides constructs such as mutexes, semaphores, and barriers.
Each tasklet is naturally able to follow its own instruction flow within the program. The \emph{me()} method, which can be called within the DPU application, returns a unique value, an integer between 0 and 23, corresponding to the \emph{tasklet} that invokes it. Based on prior considerations regarding DPU pipelining, it is recommended to use at least eleven \emph{tasklets} per DPU to efficiently hide pipeline latency.

The \emph{UPMEM PIM} architecture has been widely tested in previous work~\cite{UPMEMbench}~\cite{BenchRealPIM_UPMEM}, but the following statements can summarize the most significant characteristics:
\begin{itemize}
\item  The pipeline results completely full when 11 (or more) tasklets are executed in parallel and, in this scenario, one instruction is retired at every cycle~\cite{UPMEMbench}. It follows that DPU programs should use at least 11 tasklets in order to maximize the throughput.
\item Each DPU natively supports only bit-wise operations and addition/subtraction operations between 32-bit integers. In contrast, more complex operations (e.g., multiplication, division) or other data types (e.g., floating point) are emulated via runtime libraries~\cite{PrimerPIM}. This implies that the most well-suited workloads for the UPMEM PIM architecture use no arithmetic operations or use only simple ones~\cite{UPMEMbench}.
\item Different DPUs cannot communicate directly with each other, but information exchange among DPUs can be implemented with CPU-to-DPU and DPU-to-CPU data transfers (i.e., through a narrow memory bus)~\cite{RealPIM}. It follows that inter-DPU communication should be avoided when possible.
\end{itemize}

\section{Related Work} \label{sec:relatedwork}

To the best of our knowledge, this work is the first to implement and rigorously evaluate \emph{SHA-256}, and to conduct a comprehensive cryptographic performance study (\emph{AES-128} and \emph{SHA-256}) on a real-world PIM system. Several prior works have explored diverse algorithms and applications on the UPMEM PIM architecture, including compilers and programming models~\cite{p7,p57}, libraries~\cite{p18,p27}, simulation frameworks~\cite{p24,p26}, bioinformatics~\cite{p8,p13,p35,p36}, security~\cite{p28,he}, data analytics and databases~\cite{p2,p3,p4,p29,p38}, as well as machine learning training and inference~\cite{p11,p58,p59,p21,p60,p33,p61,p56}.

Joel Nider \emph{et al.}~\cite{PIMencryptionSASHA} investigated the implementation of \emph{AES-128} on the UPMEM architecture. Their study compared the throughput of various PIM-enabled \emph{AES-128} implementations with encryption performance on the UPMEM host CPU using the native \emph{AES\_NI} instruction set. The study concluded that the superior performance of the host CPU with \emph{AES-NI} acceleration demonstrates the inability of general-purpose DPUs to compete with specialized instruction sets. It is important to note, however, that this conclusion was based on the specific UPMEM machine version employed at the time. The algorithm was executed on a 640-DPU configuration of the UPMEM architecture, with a maximum of 512 DPUs (equivalent to 8 ranks) utilized in practice. In contrast, more recent UPMEM architectures now support up to 40 ranks, allowing the simultaneous allocation of significantly more DPUs. This expanded hardware capability may shift the comparative performance landscape and warrants a re-examination of cryptographic workloads on modern UPMEM platforms.

In a different cryptographic domain, Gupta \emph{et al.}~\cite{he} presented the first end-to-end implementation and characterization of homomorphic encryption (HE) operations on the UPMEM PIM system. Their study demonstrated that PIM can effectively accelerate memory-bound HE kernels by exploiting the large-scale parallelism. However, compute-intensive and communication-heavy kernels are limited by the lack of wide-precision arithmetic units and inter-DPU communication support, which significantly constrains overall performance.

Beyond UPMEM, several \emph{PIM} architectures have been proposed over the years with a focus on encryption and data security. One of the earliest works in this domain is \emph{CRYPTO MEMORY}~\cite{CRYPTOMEMORY}, introduced in 2004, which presented a 32\,Kbit dual-port SRAM capable of self-encrypting messages of size $16 \times 128$ bits. The concept of self-encrypting memory gained further attention with the emergence of non-volatile memory (NVM) as an alternative to traditional DRAM. Because NVM can retain data for extended periods, it introduces new security vulnerabilities. To address this, the \emph{AIM}~\cite{AES_AIM} architecture proposed an auto-encryptable, NVM-based device designed to protect sensitive data in the event of physical theft. Similarly, \emph{IMCRYPTO}~\cite{imcrypto} introduced an in-memory fabric that functions as a last-level cache reserved for secure data. This architecture exploits analog signal characteristics to accelerate basic \emph{AES} operations. Additionally, it integrates a RISC-V-based controller within the memory chips to manage the entire encryption process. Unlike fixed-function designs, the reprogrammable controller makes IMCRYPTO a more flexible and generalizable platform.

While prior works have demonstrated promising encryption-focused PIM architectures, most of these efforts have remained at the simulation or prototype stage, without resulting in commercially available systems. Moreover, no previous study has presented a comprehensive, real-system evaluation of \emph{SHA-256} on PIM hardware. Our work addresses this gap by implementing and evaluating both \emph{AES-128} and \emph{SHA-256} on the UPMEM PIM system, providing the first real-world performance and design insights for cryptographic workloads on PIM platforms. Furthermore, by leveraging the increased parallelism of newer UPMEM architectures, we revisit prior assumptions and reassess the competitiveness of PIM for cryptographic acceleration.

\section{CryptoPIM}\label{sec:cryptoPIM}
%algorithm is missing
In this section, a possible PIM-enabled software implementation of \emph{AES-128} and \emph{SHA-256} is presented.
As mentioned, AES is composed of many simple operations, making it a good candidate for the UPMEM architecture.
Moreover, since the most well-suited workloads for the UPMEM PIM architecture use no arithmetic operations or use only simple operations~\cite{BenchRealPIM_UPMEM}, some expedients can be adopted to further reduce the number of arithmetic calculations required by AES on the UPMEM architecture. As an example, in the \emph{MixColumns} step, every state array column is multiplied (i.e., finite field multiplied) by a fixed 4x4-byte matrix containing only 0x01, 0x02, and 0x03 values. The step can be sped up by pre-computing the results of the finite field multiplication between every byte in the range 0x00-0xFF and the fixed matrix coefficients and saving them in a look-up table~\cite{imcrypto}.
A Look-up-table can also be used in the key expansion process in order to store pre-calculated the ten round coefficients \emph{RC}. Since the most well-suited workloads for the UPMEM architecture require little or no communication across DPUs~\cite {BenchRealPIM_UPMEM}, the algorithm is developed so that each DRAM Processing Unit encrypts a different state array in parallel with the other DPUs.

\begin{algorithm}[t]
\caption{Host CPU pseudocode with a single rank}
\label{hostCPUalgorithm}
\textbf{Step 1: Algorithm initialization}

\SetAlgoLined
$K_s \gets \textsc{KeyExpansion}(K);$  \tcp{AES key expansion}
$allocate\_DPUs(M);$
$broadcast\_to\_DPUs(K_s);$
\BlankLine
\textbf{Step 2: Data transfer to DPUs}

\tcp{Iteration over DPUs}   
    \For{$j\gets 0$ \KwTo $M-1$}
    {
    $prepare\_transfer\_to(DPU[j],src\_buffer[j])$\;
    }
    $start\_sync\_transfer();$
\BlankLine
\textbf{Step 3: Launch DPU execution} \\
$launch\_sync\_execution(ranks);$
\BlankLine
\textbf{Step 4: Retrieving data from DPUs} \\
    \For{$j\gets 0$ \KwTo $M-1$}{$prepare\_transfer\_from(DPU[j],dest\_buffer[j])$\;
    }
    $start\_sync\_transfer()$\;
\end{algorithm}

In the DPU program, each plaintext block is stored in a \emph{buffer}, while another \emph{cryptedBuffer} is used to store encrypted blocks. Both these arrays are placed in the MRAM since they must be accessible by both the DPU and the host CPU.
Before performing any operation, data must be transferred from the MRAM to a \emph{cache} in the \emph{Working RAM} of the DPU via the \emph{mram\_read} instruction. In this scenario, two main alternatives take place: the \emph{cache} can be declared as a global array inside a DPU (using the \emph{\_\_host} attribute in the DPU application) or as a local variable inside the main method. While in the first case all tasklets of the DPU share the same cache array (on which read and write operations are not tasklets-safe), following the second way, every tasklet has its own cache array. 

In both cases, each tasklet performs the same number of MRAM read accesses, and for each memory access (\emph{CACHE\_SIZE / NR\_TASKLETS}) and (\emph{CACHE\_SIZE}) bytes are read, respectively in the first and in the second implementation. Hence, it would be better to create a local cache for each tasklet if possible, since, according to previous work~\cite{UPMEMbench}~\cite{BenchRealPIM_UPMEM}, MRAM latency has a fixed access cost and does not increase linearly with transferred data size if this is small.

Almost the same considerations can be applied to the SHA-256 algorithm. Contrary to AES-128, the hashing of a single message using SHA-256 is not easily parallelizable since the hash of every block of 64 bytes in which the original message is divided requires the knowledge of the intermediate hash values from all the previous blocks. In this scenario, the hashing process of a single message can be parallelized neither by distributing hashing among different DPUs nor among different tasklets inside a single DPU. For this reason, during this work, the hashing of \emph{1024} different messages, each of size \emph{32 kb}, is considered, for a total \emph{32 MB} of data hashed.

\begin{algorithm}[t]
\caption{Modification to host CPU pseudocode to exploit multiple ranks}
\label{hostCPUmultipleRanks}
\textbf{Step 2: Data transfer to DPUs}

    \For{$i \gets 0$ \KwTo $N-1$}{
    \tcp{Iteration over ranks}   
        \For{$j\gets 0$ \KwTo $M-1$}
        {
        \tcp{Iteration over DPUs of $i^{th}$ rank}   $prepare\_transfer\_to\_DPU(src\_buffer[i,j])$\;
        }
        \tcp{Start asynchronous transfer toward $i^{th}$ rank}
        $start\_async\_transfer(rank[i]);$
    }
$rank\_sync();$ \tcp{Wait for all transfers to finish}
\end{algorithm}

\subsection{Algorithms}
A pseudo-code representation of the algorithm is provided in Algorithm~\ref{hostCPUalgorithm}. Before initiating the encryption process, the ten round keys must be derived from the original 128-bit key. Since the computation of each byte in the expanded key depends on the value of the immediately preceding byte, the key expansion cannot be parallelized across multiple tasklets or DPUs. Consequently, this procedure is executed on the host CPU, which subsequently broadcasts the fully expanded key to all allocated DPUs (\textbf{STEP 1}). In the subsequent stages, the primary objective of the program is to encrypt an 8~MB buffer of randomly generated bytes, stored in an array within the host program. Initially, the buffer is partitioned into \emph{NUMBER\_DPU} blocks of equal size. These blocks are then assigned to DPUs in a sequential, linear manner, after which data transfers to all DPUs are performed in parallel (\textbf{STEP 2}). Once all transfers have been completed, the DPUs are launched simultaneously, and the host program remains blocked until the last DPU finishes its assigned task (\textbf{STEP 3}). Finally, in a process similar to the one described above, data is retrieved from each DPU back to the host system RAM through an iteration over all DPUs (\textbf{STEP 4}).

Algorithm~\ref{hostCPUmultipleRanks} presents a modified version of Algorithm~\ref{hostCPUalgorithm}, in which Step 2 (and Step 4, although not shown in the pseudo-code) is implemented using two nested iterations: an outer loop indexed by \textit{i}, which iterates over all allocated ranks, and an inner loop indexed by \textit{j}, which iterates over all DPUs within the $i^{th}$ rank. Once the data destined for a specific rank have been prepared, the transfers to all DPUs within that rank are initiated asynchronously. Meanwhile, the host CPU proceeds with preparing the data for the next rank.

\section{Architecture setup}\label{sec:setup}

A from-scratch PIM implementation of \emph{AES-128} and \emph{SHA-256} was developed using the UPMEM SDK 2023.2.0. This SDK offers a comprehensive toolchain, including C and Rust compilers based on LLVM 12~\cite{UPMEMsdk}, as well as a UPMEM machine simulator. These tools enable the management of DPUs and the execution of PIM-enabled programs on a standard computer. While the simulator can be used to test and verify code correctness, it does not accurately replicate the actual execution time of programs on a real UPMEM machine. Consequently, the measured performance depends on the architecture of the host system running the simulator.

The benchmarks presented in the following sections were obtained by running the algorithms on a UPMEM machine equipped with 20 PIM DIMMs, each containing 128 DPUs (i.e., two ranks) operating at a frequency of 450~MHz~\cite{UPMEMtechnology}. This setup provides a total of 160~GB of PIM-enabled memory and a memory bandwidth of 2.56~TB/s. The UPMEM non-PIM architecture consists of an Intel Xeon Silver 4215 (the \emph{host CPU}) paired with 4 × 64 GB of standard 2666~MHz DDR4 DRAM.

\section{Experimental Results}\label{sec:results}
%\flv{ADD sections for the setup which describe hw characteristics and software stack including version}

This section presents the benchmark results of both algorithms executed on a real UPMEM architecture, following the methodology established by the \emph{PrIM benchmarks}~\cite{BenchRealPIM_UPMEM}. The evaluation begins with a strong scaling experiment, where the number of tasklets is varied within a single DPU. Next, we scale the number of DPUs within a single rank. Subsequently, weak scaling is examined across the multiple ranks. Finally, we evaluate scalability beyond a single rank by exploiting both asynchronous rank execution and asynchronous rank transfer approaches.

\begin{figure}[t]

   \begin{subfigure}[h]{0.49\textwidth}
     \includegraphics[width=\linewidth]{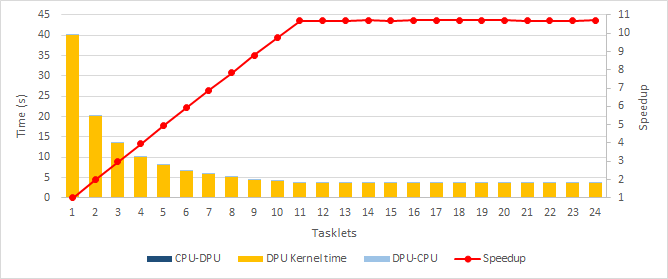}
     \caption{AES-128 tasklets scaling}\label{fig:AES128_Scaling_Tasklets}
   \end{subfigure}
   \hfill
   \begin{subfigure}[h]{0.49\textwidth}
     \includegraphics[width=\linewidth]{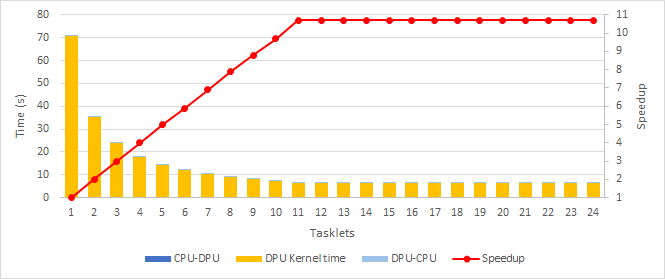}
     \caption{SHA-256 tasklets scaling}\label{fig:SHA256_Scaling_Tasklets}
   \end{subfigure}
   \caption{(a) AES-128 and (b) SHA-256 tasklets scaling. Speedups are normalized to the performance obtained using one tasklet.}
   \label{fig:taskletsScaling}
\end{figure}

\begin{figure}[t]
   \begin{subfigure}[h]{0.49\textwidth}
     \includegraphics[width=\linewidth]{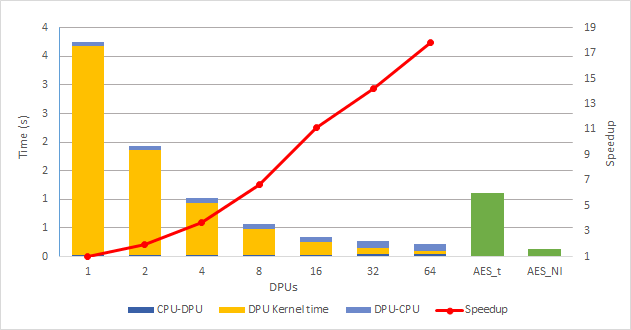}
     \caption{AES-128 strong scaling}\label{fig:AES128_DPUscaling}
   \end{subfigure}
   \hfill
   \begin{subfigure}[h]{0.47\textwidth}
     \includegraphics[width=\linewidth]{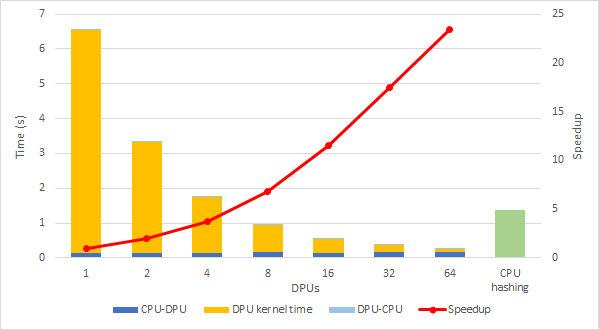}
     \caption{SHA-256 strong scaling}\label{fig:SHA256_Scaling_DPUs}
   \end{subfigure}
   \caption{(a) AES-128 and (b) SHA-256 strong scaling. Each DPU uses 16 tasklets, speedups are normalized to the performance obtained using one DPU.}
   \label{fig:strongScaling}
\end{figure}

The goal of the tasklets scaling is to check how the algorithm performs when solving a problem of fixed size, while changing the number of tasklets inside one allocated DPU. The results of the tasklets scaling, shown in Figure~\ref{fig:taskletsScaling}, match what was expected. Since each tasklet operates independently both in terms of data transfers and algorithm core functions, the speedups normalized to the performance of a single tasklet scale almost linearly with the number of tasklets up to 11. This behavior is due to the DPU pipeline, which saturates at 11 tasklets. Adding more hardware threads beyond this point does not yield additional throughput improvements. Notably, using more than 11 tasklets achieves a speedup exceeding 10$\times$ compared to the single tasklet execution. Furthermore, since the entire host buffer is transferred to the single allocated DPU, the CPU-to-DPU and DPU-to-CPU transfer times remain constant and negligible regardless of the number of tasklets.

\obs{The algorithm performance scales nearly linearly with the number of tasklets up to 11, after which the DPU pipeline saturates, and adding more tasklets does not improve performance.
}

\begin{figure}[t]
   \begin{subfigure}[h]{0.49\textwidth}
     \includegraphics[width=\linewidth]{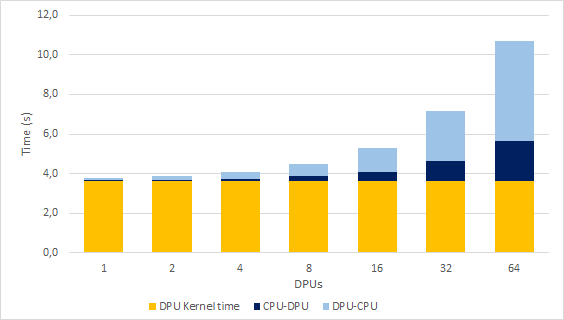}
     \caption{AES-128 weak scaling}\label{fig:AES128_WeakScaling}
   \end{subfigure}
   \hfill
   \begin{subfigure}[h]{0.47\textwidth}
     \includegraphics[width=\linewidth]{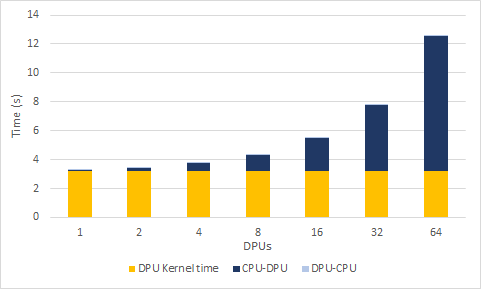}
     \caption{SHA-256 weak scaling}\label{fig:SHA256_Weak}
   \end{subfigure}
   \caption{(a) AES-128 and (b) SHA-256 weak scaling. Each DPU uses 16
tasklets.}
   \label{fig:weakScaling}
\end{figure}

Then, for both algorithms, the transfer and execution times are evaluated and reported in relation to the number of DPUs within a single rank, as shown in Figure~\ref{fig:strongScaling}. Inside each DPU, the number of tasklets is chosen based on the configuration that yielded the best performance in Figure~\ref{fig:taskletsScaling}. Although any number greater than 10 can be used, 16 tasklets are typically preferred, as this choice tends to offer a good performance. As expected, since each DPU operates independently and no inter-DPU communication is required, the DPU kernel times scale almost linearly with the number of DPUs. For comparison, Figure~\ref{fig:strongScaling} shows that the encryption and hashing of the same amount of data is performed on the host CPU of the UPMEM machine. The \emph{AES\_t} and \emph{CPU hashing} columns show the execution times for encryption and hashing on the host CPU using the same AES-128 and SHA-256 implementations deployed on the DPUs. Furthermore, since the UPMEM machine uses an \emph{INTEL Xeon Silver 4215} CPU as its host processor~\cite{UPMEMtechnology}, the host application can take advantage of the \emph{AES\_NI} instruction. This is part of an instruction set available on most of the Intel CPUs, which provides a native hardware implementation for AES encryption, significantly improving performance.

\obs{The execution times of both algorithms scale almost linearly with the number of DPUs since each DPU operates independently without inter-DPU communication, and 16 tasklets per DPU are typically preferred for good performance.
}

%WEAK SCALING

Finally, a weak scaling benchmark was conducted, in which both execution and transfer times were evaluated by keeping the problem size constant for each DPU. The results shown in Figure~\ref{fig:weakScaling} align with the ideal weak scaling scenario, where the execution time (i.e., DPU kernel time) remains constant regardless of the number of DPUs~\cite{PrimerPIM}. In contrast, the data transfer times increase linearly with the number of DPUs, as the total size of data to be transferred grows accordingly. Figure~\ref{fig:weakScaling} also reveals a strong asymmetry between CPU-to-DPU and DPU-to-CPU transfer times. This asymmetry highlights the differing performance characteristics of the two data paths. Furthermore, as shown in Figure~\ref{fig:SHA256_Weak}, the DPU-to-CPU transfer times are negligible. This is because the data retrieved from DPUs, namely the SHA-256 digests, have a fixed size of 32 bytes each, regardless of the input message size.

Despite the PIM approach, Figure~\ref{fig:weakScaling} demonstrates that when a large volume of data needs to be transferred from the host CPU to DPUs (or vice-versa), transfer times can become a dominant factor affecting the whole application performance. One way to mitigate this bottleneck is through the use of asynchronous and parallel CPU-DPU and DPU-CPU transfers. In the previous implementation of both \emph{AES-128} and \emph{SHA-256} host programs, these transfers were performed serially. All data transfers were completed before any DPU execution began, and the execution of all DPUs was launched simultaneously. Moreover, the DPUs execution was synchronous with the host program, meaning that the host application was suspended until the requested DPUs completed their execution or encountered an error~\cite{UPMEMsdk}. As a result, the host CPU remains idle during the PIM-based encryption or hashing.
The main idea of the asynchronous CPU-DPU transfers consists of preparing the data buffers one at a time and initiating each transfer asynchronously. This allows the host CPU to launch subsequent transfers or perform other computations while earlier transfers are still in progress. However, due to architectural constraints in the UPMEM architecture, asynchronous transfers are only supported across different \emph{ranks}, not across individual DPUs within the same rank. Therefore, to take advantage of asynchronous transfers, the application must be designed to utilize multiple ranks.

\obs{In weak scaling, execution time remains constant as the number of DPUs increases, but data transfer times grow linearly and can dominate performance; using asynchronous transfers across multiple ranks can help mitigate this bottleneck, though asynchronous transfers are not supported within a single rank.
}

\begin{figure}[t]
   \begin{subfigure}[h]{0.48\textwidth}
     \includegraphics[width=\linewidth]{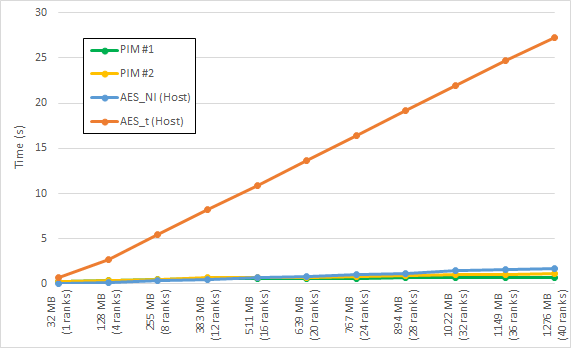}
     \caption{AES-128}\label{fig:AES128_RankScaling1}
   \end{subfigure}
   \hfill
   \begin{subfigure}[h]{0.49\textwidth}
     \includegraphics[width=\linewidth]{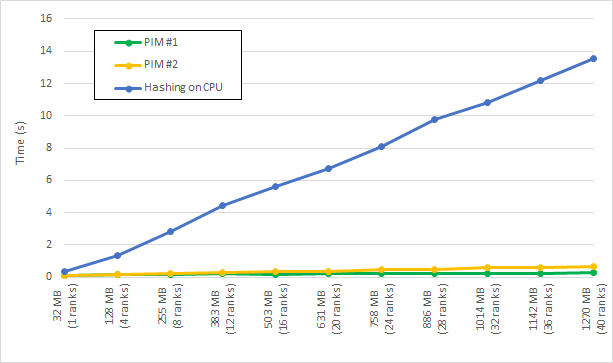}
     \caption{SHA-256}\label{fig:SHA256_Asyn_CPUcompared}
   \end{subfigure}
   \caption{(a) AES-128 and (b) SHA-256 weak scaling using up to 40 ranks. \emph{PIM \#1} refers to asynchronous rank transfers, while \emph{PIM \#2} to asynchronous rank execution. In both PIM versions, 16 tasklets are allocated for each DPU.}
   \label{fig:rankScaling1}
\end{figure}

The UPMEM SDK~\cite{UPMEMsdk} also provides the capability to perform asynchronous rank launches, allowing the host application to continue executing while launched ranks carry out their computations. In this mode, once all DPUs within a rank have been initialized, the data transfer to that rank is performed synchronously, followed by an asynchronous launch of the rank. This approach makes it possible to do computation and data preparation at the same time. The host application can prepare data transfers for the next ranks while the previous ranks are still running. However, it is important to understand that data transfer from the CPU to the DPU and the execution on the DPU cannot happen at the same time for one rank. Before a rank starts running, all its data must be fully transferred. This rule limits how much overlap there can be between data movement and computation, but using pipelining with several ranks can still improve performance.

As mentioned before, recent versions of the UPMEM machine support up to 40 ranks, providing a total of 2560 DPUs. In practice, however, the number of usable DPUs may be slightly lower due to the presence of some faulty devices~\cite{BenchRealPIM_UPMEM}).

To test how the algorithm works with more ranks, another weak scaling test was performed. In this test, the plaintext size was fixed at 32~MB per rank for both encryption and hashing. Figure~\ref{fig:rankScaling1} shows a comparison of the performance of three versions of the algorithm. The first version is a PIM implementation that uses asynchronous rank transfers (\emph{PIM \#1}). The second version is another PIM implementation that uses asynchronous rank execution (\emph{PIM \#2}). The third version is a non-PIM implementation that runs on the host CPU of the UPMEM machine~\cite{UPMEMtechnology}. Furthermore, for AES-128, a fourth version is also considered. This is a non-PIM implementation that runs on the host CPU of the UPMEM machine~\cite{UPMEMtechnology} and uses the native Intel \emph{AES\_NI} ISA extension to improve performance.

\begin{figure}[t]
   \begin{subfigure}[h]{0.48\textwidth}
     \includegraphics[width=\linewidth]{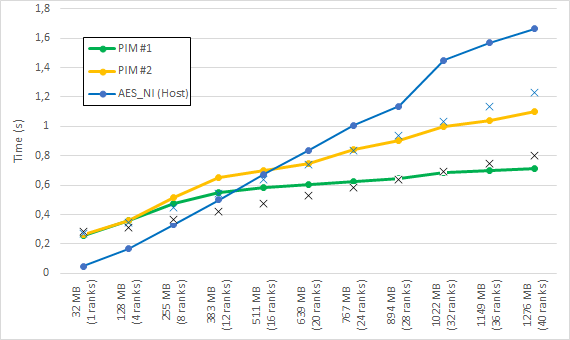}
     \caption{AES-128}\label{fig:AES128_RankScaling2}
   \end{subfigure}
   \hfill
   \begin{subfigure}[h]{0.51\textwidth}
     \includegraphics[width=\linewidth]{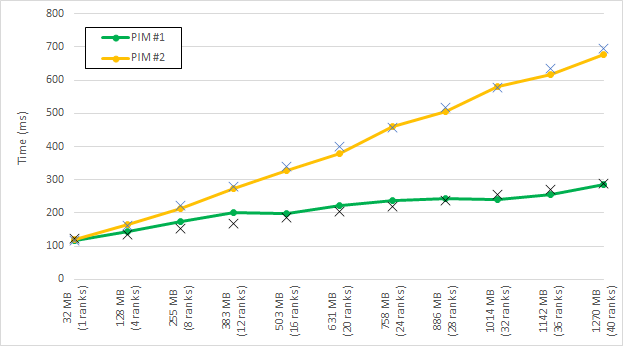}
     \caption{SHA-256}\label{fig:SHA256_Asyn_Estimated}
   \end{subfigure}
   \caption{Graphs in figure \ref{fig:rankScaling1} omitting host CPU software performance. Black and blue crosses represents \emph{PIM \#1} and \emph{PIM \#2} implementations.}
   \label{fig:rankScaling2}
\end{figure}

As expected, the time required by the software non-PIM execution is up to an order of magnitude above all others. For this reason, these two lines have been excluded from the pictures reported in Figure~\ref{fig:rankScaling2}.

\obs{Asynchronous rank launches allow overlapping computation and data preparation across ranks, improving performance through pipelining. Weak scaling tests show that PIM implementations significantly outperform non-PIM CPU versions, especially for large workloads.}

%As mentioned earlier, a previous study~\cite{PIMencryptionSASHA} has already investigated \emph{AES-128} implementation on the UPMEM architecture. Figure~\ref{fig:PIM_AES_SASHA} presents the cryptographic results from that work, comparing the throughput of various PIM-enabled versions of AES-128 with the performance achieved by encryption on the UPMEM host CPU utilizing the native \emph{AES\_NI} instruction set.

%\begin{figure}[t]
%\centering
%\includegraphics[scale=0.3]{Images/PIM_AES_SASHA.png}
%\caption{Encryption throughput of PIM AES-128 implementations %compared to a single host CPU~\cite{PIMencryptionSASHA}.}
%\label{fig:PIM_AES_SASHA}
%\end{figure}

%The conclusion of the previous study, which states that the superior performance on the host with \emph{AES-NI} acceleration confirms that general-purpose DPUs cannot compete with specialized instruction sets~\cite{PIMencryptionSASHA}, was based on the specific UPMEM machine version employed at the time. The algorithm was executed on the 640-DPUs configuration of the UPMEM architecture; as shown in Figure~\ref{fig:PIM_AES_SASHA}, a maximum of 512 DPUs (equivalent to 8 ranks) were utilized. In contrast, more recent UPMEM architectures support the simultaneous allocation of up to 40 ranks, potentially altering the comparative performance landscape.

\section{Conclusions}\label{sec:conclusions}
This paper presents the scaling of two of the most widely used cryptographic algorithms (AES-128 and SHA-256) on a real-world PIM system (i.e., UPMEM). These algorithms are well-suited to the PIM architecture for several reasons. First, in most traditional architectures, their software implementations tend to be \emph{memory-bound}. Second, both algorithms consist primarily of simple operations, such as bitwise XOR and AND. Third, they operate on bytes or 32-bit integers, thereby avoiding the need for 64-bit integers and floating-point operations, which are often not supported natively in the hardware of real-world PIM architectures. Lastly, encryption and hashing algorithms are highly parallelizable across many processing units. When these algorithms are scaled on a single rank, the results are consistent with previous work, although they still fall short of software implementations on modern CPUs. To further improve application performance, scaling across multiple ranks (sets of 64 DRAM Processing Units) is essential. For this reason, the paper explores and evaluates two execution models: asynchronous rank transfer and asynchronous rank execution.  

The results indicate that asynchronous transfers (\emph{PIM \#1}) should be preferred over asynchronous executions (\emph{PIM \#2}) when no inter-DPU communication is required. By utilizing all available ranks on the real-world PIM system, cryptographic algorithms can be effectively accelerated. These findings highlight the potential of real-world PIM systems as viable alternatives to traditional processor-centric architectures, offering a highly parallel and efficient environment particularly well-suited for memory-bound cryptographic applications.

\section*{Acknowledgments}
This work has been funded by the European Union via the European Defence Fund project ARCHYTAS under grant number 101167870. Views and opinions expressed are however those of the author(s) only and do not necessarily reflect those of the European Union or the European Commission. Neither the European Union nor the granting authority can be held responsible for them.

%\begin{acks}
%To Robert, for the bagels and explaining CMYK and color spaces.
%\end{acks}

%%
%% The next two lines define the bibliography style to be used, and
%% the bibliography file.
%\bibliographystyle{ACM-Reference-Format}
%\bibliography{sample-base}

\printbibliography

\end{document}